\documentclass[]{aastex}
\usepackage{emulateapj5}
\slugcomment{The Astrophysical Journal, in press}
\shorttitle{Planet around $\gamma$~Cephei A}
\shortauthors{Hatzes et al.}
\begin{document}

\title{A Planetary Companion to $\gamma$~Cephei A}

\author{Artie P. Hatzes}
\affil{Th\"uringer Landessternwarte, D - 07778 Tautenburg, Germany}
\email{artie@jupiter.tls-tautenburg.de}

\author{William D. Cochran, Michael Endl, Barbara McArthur, Diane B. Paulson}
\affil{McDonald Observatory and Astronomy Department,
The University of Texas at Austin, Austin, TX 78712}
\email{wdc@astro.as.utexas.edu, mike@astro.as.utexas.edu,
mca@astro.as.utexas.edu, apodis@astro.as.utexas.edu}

\author{Gordon A. H. Walker}
\affil{Physics and Astronomy Department, University of British Columbia,
Vancouver, B.C. Canada V6T 1Z4}
\email{walker@astro.ubc.ca}

\author{Bruce Campbell}
\affil{BTEC Enterprises Ltd.}

\and

\author{Stephenson Yang}
\affil{Department of Physics and Astronomy, University of Victoria,
Victoria, BC, Canada, V8W 3P6}
\email{yang@uvastro.phys.uvic.ca}

\begin{abstract}
We report on the detection of a planetary companion in orbit around the primary
star of the binary system $\gamma$~Cephei. High precision
radial velocity measurements using  4 independent data sets spanning the
time interval 1981--2002  reveal long-lived residual radial velocity
variations superimposed on the binary orbit
that are coherent in phase and amplitude
with a period or 2.48 years (906 days) and a semi-amplitude of 27.5 m\,s$^{-1}$.
We performed
a careful analysis of our Ca II H \& K S-index measurements, spectral
line bisectors, and {\it Hipparcos} photometry. We found no significant
variations in these quantities with the  906-d period. 
We also re-analyzed
the Ca II $\lambda$8662\,{\AA} measurements of Walker et al. (1992)
which showed possible periodic variations with the ``planet'' period
when first published. This analysis shows that periodic Ca II equivalent
width variations were only present during 1986.5 -- 1992 and 
absent during 1981--1986.5. Furthermore, a refined period for the
Ca II  $\lambda$8662\,{\AA} variations is
2.14 yrs, significantly less than residual radial velocity period.
The most likely explanation of the residual radial velocity variations 
is  a planetary mass companion with $M$ sin $i$ = 1.7 $M_{Jupiter}$
and an orbital semi-major axis of $a_2$ $=$ 2.13 AU.  
This supports the planet hypothesis for the residual radial velocity
variations for $\gamma$ Cep first suggested by Walker et al.
(1992). With an estimated binary orbital period of 57 years 
$\gamma$~Cep is the shortest period binary system in which an extrasolar planet has 
been found. This system may provide insights into the relationship between planetary
and binary star formation.
\end{abstract}


\section{Introduction}
\label{intro}
The first high-precision radial velocity survey for planetary companions
to nearby stars was conducted with the use of a HF gas absorption cell
on the Canada-France-Hawaii Telescope (Campbell \& Walker 1979; 
Walker et al. 1995).
The $\gamma$~Cephei system (HR~8974 = HD~222404 = HIP~116727) was one of
16 stars on the observing list of this program.
Campbell, Walker, \& Yang (1988)  reported that two of the stars on the observing list
($\gamma$~Cephei and $\chi^1$~Orionis)
were previously unknown single-lined spectroscopic binaries.
In the case of $\gamma$~Cephei, Campbell, Walker, \& Yang (1988) found evidence for radial
velocity ``bumps'' superimposed on the large amplitude binary motion.
The residuals after removing the binary orbit yielded a period of about
2.7~years and a velocity semiamplitude of about 25\,m\,s$^{-1}$.
The authors examined several possible causes of the observed radial
velocity variability, and finally concluded that the system had a ``probable
third body.''   If substantiated, this would have been the first detection of
an extrasolar planetary system.   However, Bohlender, Irwin, \& Yang (1992) classified
$\gamma$~Cep as a K0~III star, raising the possibility that the observed
radial velocity variations were due to the recently discovered
(Walker et al. 1989; Hatzes \& Cochran 1993) long-period radial velocity variability of most
K~giants.  A detailed analysis of the CFHT radial velocity data on
$\gamma$~Cep by Walker et al. (1992) showed a clear low-amplitude (27\,m\,s$^{-1}$)
signal with a 2.52~year period superimposed on the binary orbital motion of
indeterminate period.   They also found evidence of a possible variation
of the Ca~II $\lambda$8662\,{\AA} emission line index with the same 2.5 year
period, leading them to conclude that the observed low-amplitude radial
velocity variation was most likely due to K-giant variability at the period of
the stellar rotation.

Here we present new high-precision radial velocity data on the $\gamma$~Cephei
system obtained from McDonald Observatory.  When combined with the CFHT data,
we show that the 2.5-year low-amplitude radial velocity variability of
$\gamma$~Cep has remained constant for over 20 years.  A 
simultaneous orbital solution for both the binary star and the 2.5-year
low amplitude variability is computed.  We demonstrate that there is no correlation
between the low-amplitude radial velocity variations and the Mt.~Wilson 
Ca~II S-index, 
nor is there any indication of
photospheric absorption line profile variability.  These results, combined
with the current best classification of $\gamma$~Cep as a K1\,IV star (Fuhrmann
2003), all
indicate that the preferred interpretation of the data is that $\gamma$~Cep~A
has a planetary-mass companion in a 2.5~year period orbit.

\section{The Radial Velocity Data Sets}
We consider four independent sets of high precision radial velocity data for
$\gamma$~Cephei, covering the interval from 1981 through 2002.
The first of these is from the CFHT survey Campbell, Walker, \& Yang (1988),
Walker et al. (1995), with
the velocities taken from Table~1 of Walker et al. (1992).  
All of the other data
were from the McDonald Observatory Planetary Search (MOPS) Program
(Cochran \& Hatzes 2000).
Phase~I of the MOPS program used the telluric O$_2$ lines near
6300{\AA} as the velocity metric, a technique suggested by Griffin \& Griffin
(1973).  A single order of the 2.7m coud\'e spectrometer 6-foot camera 
with the echelle grating was
isolated onto a Texas Instruments (TI) $800 \times 800$ CCD at $R = 210,000$.  This system gave
15--20\,m\,s$^{-1}$ precision on stars down to about $V = 6$, but suffered from
systematic velocity errors, most likely due to prevailing atmospheric winds.
In 1992 the program switched to a temperature stabilized I$_2$ cell
(Koch \& W\"ohl 1984; Libbrecht 1988) as the velocity metric for Phase~II of
the MOPS ($R = 210,000$ for this data as well).  
This eliminated the systematic errors, and gave a routine radial
velocity precision of 15\,m\,s$^{-1}$. 
This precision was limited by the 9.6\,{\AA} bandpass of the
spectrum, and by the poor charge-transfer and readout properties of the TI CCD.
To solve these problems, and to achieve substantially improved precision, we
began Phase~III of the radial velocity program in July 1998, using the I$_2$
cell with the newly installed 2dcoud{\'e}
cross-dispersed echelle spectrograph
(Tull et al.~1994).  This instrument when used with a Tektronix 2048$\times$2048
detector  provides a nominal wavelength coverage of 
3600\,{\AA} -- 1$\mu$m at a resolving power of $R = 60,000$. The
Tektronix CCD also had significantly better charge transfer and readout
properties than the TI device.
The complete spectral coverage of 2dcoud\'e
gives us two advantages: first, we can utilize the full reference spectrum of
the I$_2$-cell for the RV determination and second, we can simultaneously
determine the stellar chromospheric emission in the cores of the
Ca\,II~H\&K lines to use as stellar chromospheric activity indicators.
We will discuss in detail the question whether there is a correlation between
the RV-results and the activity indices for $\gamma$~Cep in section \S4.

To extract the RV-information from the I$_2$ self-calibrated spectra taken 
during Phase~III we
employed our $Austral$ RV-code, which uses a Maximum Entropy Method deconvolution
in order to obtain a higher resolved stellar template spectrum and several
reconstruction algorithms which model the shape and symmetry of the
instrumental profile (IP) at the time of observation.
A detailed description of the $Austral$ code can be found in Endl, K\"urster, \& Els (2000).
The algorithm follows in general the modeling idea first outlined by
Butler et. al. (1996), and the IP reconstruction techniques by Valenti et al. 
(1995).
Using the 2dcoud{\'e} spectrometer in I$_2$-cell self-calibration mode and the
$Austral$ code for the analysis, we obtain a long term RV precision of
$5 - 15~{\rm m\,s}^{-1}$ on a routine basis for stars down to a magnitude
of $V=9.0$.

All of the MOPS data from Phases I, II and III are given in Tables 1--3.
The uncertainties quoted there for Phase II and III data are the ``internal''
errors, as represented by the rms of the individual spectral chunks about
the mean value.  We regard these as a lower limit on the actual uncertainties,
since these values do not include the effects of any residual systematic
errors that may be present.  It is difficult
to estimate internal errors for the Phase~I data because
these are not analyzed in ``chunks'' like the Phase II and III
measurements. Phase~I observations of a constant star ($\tau$ Cet)
show an rms scatter of 23 \,m\,s$^{-1}$. 
The rms scatter of the Phase I measurements
about the final orbital solution is 17.4\,m\,s$^{-1}$. We list a  slightly
higher value of 19 \,m\,s$^{-1}$ as the ``error'' of the Phase I measurements
in Table 1. This error was the most reasonable error to assign based upon
$\chi^2$ tests of the data and was the value assumed for the Phase~I
measurements in the final orbital solutions.
The true error for an individual Phase I measurement
is almost surely higher due to different signal-to-noise ratios of individual
spectra and systematic errors due to wind and temperature and pressure
changes in the earth's atmosphere.

	The RV measurements for all data sets are shown in Figure~\ref{orbit}.
A different velocity offset (see below)  had to be applied to each data
set so that they would all have the same zero point. One can clearly see
that the ``wiggles'' superimposed on the binary variations, that were first
reported by Walker et al. (1992) are still present in the final data set 
(Phase III) taken 20 years later.

	Figure~\ref{periodogram} shows a sequence of Lomb-Scargle
periodograms  (Lomb 1976, Scargle 1982) of the RV measurements
after subtraction of the velocity variations
due to the binary companion (see below). The top panel is only
for the CFHT data. The central panel is for the  CFHT + McDonald data,
excluding the Phase III data. The lower panel is the periodogram
for all of the RV measurements. The increase in power at a period of
$\approx$ 2.5 yrs with the addition
of each data set is the first indication that  these residual
RV variations are long-lived and coherent. (The false alarm
probability of the peak for the full dataset is $\approx$ 10$^{-20}$).

\section{Orbital Solutions}

The program {\it GaussFit}  (Jefferys et al. 1988; McArthur et al. 1994) was
used to fit simultaneously all orbital parameters, both for the stellar and the
presumed sub-stellar companion, using nonlinear least squares
with robust estimation. Because each instrument
produces relative radial velocities with their own arbitrary zero point,
the individual velocity offsets were a free parameter in the least
squares solution. Table~\ref{offsets} gives the velocity offsets
of the different data sets. Subtracting these velocity offsets from
an individual data sets will place them all on the same radial 
velocity scale.
The line in
Figure~\ref{orbit} shows the combined orbital solution to $\gamma$~Cep.
Figure~\ref{residuals} shows the ``planet only'' orbital solution after
subtracting the contribution of the binary orbit to the RV measurements.

	Table~\ref{binorbit}
 lists the orbital parameters for the  the stellar companion.
The errors listed are the correlated errors that are produced by
{\it GaussFit}.
The largest error occurs in the period since our observations span 
about 20 years, or about one-third of the binary orbit. Note that the
reduced $\chi^2$ of the binary ``only'' orbital solution is rather
large (4.36) indicating the presence of additional RV variations.

	Griffin et al. (2002) presented an orbital solution for the stellar
companion to $\gamma$ Cep. They combined 
radial velocity measurements
spanning over 100 years taken at five
observatories.  These measurements included the CFHT data as
well as some of our McDonald Observatory measurements (read from an
enlarged copy of  a published graph!). The parameters
of their orbital solution is also listed in Table~\ref{binorbit}.
In spite of the more limited time span of our observations our orbital
solution agrees quite well with the Griffin et al. orbit.

	Table~\ref{planorbit} lists the {\it GaussFit}
 orbital parameters for the sub-stellar companion. Also listed
is the rms scatter of the individual datasets 
about the combined orbital solution.
The period of 2.48 years and 
semi-amplitude of 27.5 m\,s$^{-1}$ are consistent with the values
found by Walker et al. (1992) using only the CFHT data set. 
Note that the reduced  $\chi^2$ is significantly lower ( $\chi^2$ =
1.47) when including the planet in the orbital solution.
Fuhrmann (2003) estimates the primary mass of $\gamma$~Cep as
$M$ $=$ 1.59 $\pm$ 0.12 $M_\odot$. This results in a planetary mass
of $M_p$ $sin$\,$i$ $=$ 1.7   $\pm$ 0.4 $M_{Jupiter}$ and an orbital
semimajor axis of 2.13 AU. 

Figure~\ref{phase}
shows the phase diagram of the individual data sets. The phase, amplitude, and
overall shape of the RV curves hold up remarkably well over a 20 year time span,
again evidence for long-lived and coherent variations that are consistent
with the presence of a  planetary companion.

\section{The Nature of the RV Variations}

	The question naturally arises as to whether some phenomena
instrinsic to the star
(spots, convective shifts, pulsations) are responsible for the 2.48 yr
period. 
After all, $\gamma$~Cep is most likely a 
subgiant (see below), a class of stars for which magnetic activity, surface structure,
convection, and pulsations are poorly known. Moreover, Walker et al. (1992)
did find evidence for weak periodic variations in the 
equivalent
width of the Ca II $\lambda$8662\,{\AA} line with the same period
as the planet.  In spite of the intrinsic variability of giant stars 
planetary companions have been found around other
giant stars (Frink et al 2002; Setiawan et al. 2003) and one subgiant
(Butler et al. 2001).
So planets around giant evolved stars is not an unreasonable hypothesis.
Before concluding that the planet
hypothesis is the most likely explanation for the residual
RV variations we must demonstrate that $\gamma$~Cep does not exhibit any significant
variations with the planet period in other quantities.
Here we examine whether $\gamma$~Cep also has
spectral and photometric variations.

\subsection{Bisector Analysis}

	The spectral line shapes can also provide evidence in support 
of the planetary hypothesis. Surface features (Hatzes 2002)
or nonradial pulsations  (Hatzes 1996, Brown et al. 1998)
should produce changes in the spectral line shapes with the same 
period as the RV variations. A convenient means of measuring the 
asymmetry of a spectral line is the line bisector, or the line 
segments connecting the midpoints of the spectral line from the core 
to the continuum.  Line bisector studies have been used to 
establish the planetary nature of 51 Peg (Hatzes, Cochran, \&
Johns-Krull 1997; Hatzes, Cochran, \& Bakker 1998a,b) 
and the starspot nature of the planet-like
RV variations in HD 166435 (Queloz et al. 2001).

	The McDonald Observatory Phase I 
observations taken in the 6300 {\AA}  region
provide an excellent data set for studying possible line profile variations in 
$\gamma$~Cep. The data has very high spectral resolution ($R$ = 210,000) and 
there are several strong spectral lines that were velocity shifted clear of
the telluric features because of the Earth's barycentric velocity.
The
Fe I $\lambda$6301.5 {\AA} feature was free of telluric lines over the entire
data set, Fe I  $\lambda$6297.8  for 30  observations, and  
Fe I  $\lambda$6302.5  {\AA} for 8 observations.
Both the bisector velocity span and curvature were measured for these lines. 
The velocity span is defined as the velocity difference between two 
arbitrary points on the line bisector while the curvature is the 
difference between the velocity span of the top half of the bisector minus 
the velocity span of the lower half of the bisector.

A 5th order polynomial was fit to each measured line bisector
and the  velocity span
points taken at 0.4 and 0.8 of the continuum. (In measuring the line
bisector one should avoid both the core 
and the continuum where the errors in the bisector
measurement become large.) 
The additional point required for the
curvature measurements was taken at 0.6 of the continuum value. The mean value 
of the bisector span and curvature for each line was then subtracted from 
the individual measurements and the residual span and velocity measurements 
were then averaged, weighted by the rms scatter of the 
bisector measurements for each spectral line.

	Figure~\ref{bisector} shows the phase-binned averages ($\Delta\phi$
$\approx$ 0.05)
of the  bisector span and 
curvature variations for $\gamma$~Cep phased to the planet orbital period. 
The short-dashed horizontal line gives the zero-point reference and
the long-dashed lines show the velocity extrema of the residual 
RV variations due to the planetary companion. There are no obvious phase 
variations of the bisector quantities with the planet period. The 
least squares sine fit to the bisector
quantities assuming a period of 906 days yields amplitudes of 4.8
$\pm$ 4.4 m\,s$^{-1}$ and  2.4 $\pm$ 4.5 m\,s$^{-1}$
for the bisector and span variations, respectively. These amplitudes are
consistent and are significantly less than 
the observed 27 m\,s$^{-1}$ radial velocity variations of 
the planet.

	The McDonald Phase I data had the largest errors of all
the data sets and
the phase variations
of these show the least evidence
for the 906 day period (Fig.~\ref{phase}). One could argue 
that there are
no residual RV variations with which to correlate
to the bisector measurements. However, the Phase I data is still 
consistent with the presence of the planet signal.
Not only is this signal 
present in the Phase I 
and II data alone, but the power increases by almost a factor of two 
over the individual periodograms when the two data sets are combined.
Although not obvious to the eye, the 906-day period is still present in the 
Phase I data. On the other hand,
the averaged, phase-binned
bisector measurements have variations
significantly less than the 906-day RV amplitude and we believe this
adds additional evidence (along with the photometric and Ca II
analysis presented below) in support of the planet hypothesis.
{\it If}  the bisector variations had scatter comparable to the
RV amplitude, then one could at least make
a plausible argument in favor of possible bisector variability. 
Figure~\ref{bisector}  excludes that.
	
\subsection{Photometric Variations}

	The {\it Hipparcos} mission (Perryman 1997) took  high precision
photometric measurements of $\gamma$~Cep  contemporaneously with
the measurements used for our RV study. If cool spots on the
stellar surface, or some
form of stellar pulsations were causing the residual RV variations, then this 
should be evident in the {\it Hipparcos} photometry.

	Figure~\ref{photometry} shows the {\it Hipparcos} photometry taken
from 1989.9 to 1993.2 phased to 
the planet orbital period. Crosses represent individual measurements while
solid points represent phase-binned averages ($\Delta \phi$ $<$ 0.05).
Error bars represent the rms scatter of the measurements used for each bin.
There are no obvious photometric variations with the planet
period as confirmed by a periodogram analysis of the daily averages of the
{\it Hipparcos} photometry (Figure~\ref{photft}).
 A least squares sine fit
to the phased photometric
data yields an amplitude of $\Delta V$ =
0.001  $\pm$ 0.0009 mag for any possible photometric variations.
(A fit to the phase-binned measurements yields  $\Delta V$ = 0.0003.)
The lack of photometric
variations in $\gamma$~Cep is also consistent with the planet hypothesis
for the residual RV variations for this star.

\subsection{Ca II Variations}
\subsubsection{S-index Measurements}

Stellar activity in
$\gamma$~Cep could induce significant periodic
centroid shifts in photospheric absorption lines
which could be confused for
perturbations made by planetary companions (Saar \& Donahue 1997;
Saar \& Fischer 2000, Queloz et al. 2001). For the Phase III data
an instrumental setup was chosen so as to include
the Ca {\sc ii} H and K
lines on the detector.

To measure stellar chromospheric activity, the Mt. Wilson $S$ index was
adopted.
This index is defined
(e.g. Baliunas et al. 1992) as a quantity proportional to the sum of
the flux in 1\,{\AA} FWHM triangular
bandpasses centered on the Ca II H and K
lines divided by the sum of the flux in 20{\AA} bandpasses in the continuum at
 3901 and 4001{\AA} (Soderblom, Duncan, \& Johnson 1991).

The four quantities to be measured (the
two calcium line core fluxes plus
the two continuum bandpass fluxes) are spread across 3 echelle spectral orders
which overlap by several {\AA}.
In order to be consistent with activity monitoring with our other programs 
(Paulson et al. 2002),
we did not use measurements in the bluemost order, i.e. the continuum region
centered on 3901{\AA}.
In addition, we did not measure the Ca II  H line (at 3968.47{\AA}) because
the wings
of strong Balmer H$\epsilon$ feature (at 3970.07{\AA}) are within the measured
Ca {\sc ii}
bandpass. This can adversely affect the measurement of the Ca {\sc ii} 
H line flux.
Therefore, we have defined an index
$S_{\rm McD}$ which is the
ratio of the flux in a 1\,{\AA} triangular bandpass centered on the Ca {\sc ii} K
line to the
flux in a 20\,{\AA} bandpass centered on the redward continuum at 4001\,~{\AA}.
Thirty of our program stars (for the McDonald Observatory Planet Search) have
previously been measured as part of the Mt.
Wilson survey (Baliunas et al. 1995; Duncan et al. 1991). 
We find a linear relationship between our measurements and those of the
Mt. Wilson survey of the form:
\begin{equation}
S_{\rm Mt. Wilson} = 0.038 (\pm 0.006) + 1.069 (\pm 0.040) \times S_{\rm McD}
\end{equation}
\noindent
Both the slope and the intercept of the formal fit are within 1 $\sigma$ of
$S_{\rm McD}=S_{\rm Mt. Wilson}$. We are thus able to transform our data
into a standard Mt. Wilson $S$ index scale using
Equation (1). 

The periodogram of the S-index measurements (Figure~\ref{sindexft})
shows no
significant power at the orbital frequency of the planet. The
false alarm probability of the highest peak, assessed using a bootstrap
randomization process (Murdoch et al. 1993; K\"urster et al. 1997),
is 50\%.  The lack of variability at the planet period is substantiated 
by phasing these variations to the 906 day period (Figure~\ref{sindex}). 
Clearly no
significant sinusoidal 
variations in this chromospheric index are present. The transformed
Ca II K S-index for $\gamma$~Cep  is $-$5.3 which implies a level of magnetic
activity less than that of the Sun.

\subsubsection{Analysis of the Ca II $\lambda$8662\,{\AA} data from Walker et al.
(1992)}

	A conclusion regarding the planetary nature of the residual RV 
variations of $\gamma$~Cep would not be complete without a discussion of the 
Ca II variations found by Walker et al (1992). Although the  McDonald 
S-index measurements do not show evidence for rotational modulation, 
Ca II $\lambda$8662\,{\AA} equivalent width measurements ($W_\lambda$) of 
Walker et al. (1992) did show 
a hint of sinusoidal variations when phased to the planetary orbital period. 
Although this on its own does not completely refute the existence 
of a planet, having Ca II variations with the same period would 
cast more doubt on this hypothesis.
For these reasons we
made a careful examination the significance of the Ca II variations
reported by Walker et al. (1992).

	Figure~\ref{cacorr} shows the correlation between
the RV and the changes in the Ca II $\lambda$8662\,{\AA} equivalent
width. The two quantities show no obvious correlations. The 
correlation coefficient, $r$, is only 0.08 and the probability that the two
quantities are not correlated is 0.52. However, there are 3 obvious
outliers in the figure ($|\Delta W_\lambda| > $ 3\%).
Eliminating these 3 points increases the correlation coefficient to
$r$ = 0.23 with a probability of 0.08 that the
quantities are uncorrelated. However, this is still not a strong correlation.

The
McDonald S-index data also do not seem to be correlated to the RV
measurements.
Figure~\ref{mcdcorr}  shows the correlation between the 
S-index and radial velocity measurements from the McDonald data.
The correlation coefficient in this case is $r$ = $-$0.3 and the probability
that the two quantities are uncorrelated is 0.13.

	Figure~\ref{cfhtcaperiod} shows the Lomb-Scargle periodogram of the
variations in the  Ca II $\lambda$8662\,{\AA}  equivalent width measurements 
of Walker et al. (1992). 
(There was a  clear outlier in the Walker et al. Ca II data
with $\Delta EW$ = $-$4.3. This was 
also evident in Fig. 3 of Walker et al. (1992). This data point was
eliminated before performing the periodogram analysis.)
Although the most power occurs
at a periods of $\approx$ 15 days (top panel), there appears to be 
significant power at the planet period 
in the expanded scale in the
lower panel (the vertical line indicates the location of the planet
RV frequency).  This was the basis for the Walker et al. favoring 
rotational modulation as the cause of the RV variations.

A more detailed examination of the CFHT
Ca II $\lambda$8662\,{\AA} equivalent measurements show that the 
variation in these are not long-lived and may have no relation 
to the observed 906-day RV period.
We divided the CFHT Ca II data into two data sets, one spanning
1981--1986.5 and the second 1986.5--1992. The division of the data set was taken
so as to maximize the power in the long period variations
found in the second set. 
(Our results do not change substantially
if we divide the data set into two equal numbered points.) 
This resulted in 21 data points in the first
data set and 29 in the second set. 

Figure~\ref{casplit} shows the periodograms of these two data sets.
The 1981-1986.5 set shows no power in the frequency interval $0<$ $\nu$ $<$ 
0.01 c\,d$^{-1}$. The 1986.5--1992 data set shows significant power, but at
a frequency corresponding to a period of 781 $\pm$ 116 days, significantly
less than the planet orbital period. Extending the periodogram to higher 
frequencies shows 
that this is the highest peak out  to the Nyquist frequency,
in contrast to the case for the full data set where the low frequency
feature was the second highest peak. A periodogram analysis
of the full Ca II data set lowers the Lomb-Scargle power near $\nu$ = 0.011
c\,d$^{-1}$
from about 9.0 to 6.7 and increases the period slightly. 
Unlike the case for the RV data, increasing
the number of measurements does not increase the power in
the periodogram of the feature of interest. 
Clearly this signal is not long-lived.

	The statistical significance of this signal was examined using 
the bootstrap randomization technique.
The Ca II $\lambda$8662\,{\AA} equivalent width measurements over
the time span 1986.5 -- 1992 were 
randomly shuffled keeping the observed times fixed. 
A periodogram was then computed
for each  ``random'' data set. The fraction of 
the periodograms having power higher
than the data periodogram in the range 0.0005 $<$ $\nu$ $<$ 0.01 c\,d$^{-1}$
is the false alarm probability that noise would create the detected
signal. After  10$^{5}$ shuffles there  was no instance of a random
periodogram having power higher than that found in the data periodogram.
The false alarm probability is thus $<$ 10$^{-5}$. This signal, in spite of
only being present in the data for the last half of the data set, is
highly significant.

	The period of the Ca II $\lambda$8662\,{\AA} measurements is 
significantly less, by 1$\sigma$, from the residual RV period.
Figure~\ref{2phase} shows the phase diagrams of the Ca II  $\lambda$8662\,{\AA}
measurements. The top panel are the measurements during 1986.5-1992
phased to the 781 day period found in the Ca II data. The
central panel is the same data phased to the 906 day residual
RV period. Although there are slight sinusoidal variations when phased to the
906 day period, phasing
to the 781 day period produces significantly 
less scatter. By comparison, the lower 
panel shows the Ca II measurements from 1981 -- 1986.5
phased to the 906 day period. Crosses are individual measurements whereas
solid points represent phase-binned averages (the error bar represents
the scatter of the data used for each phase bin).  There are no significant
sinusoidal variations in the 1981 -- 1986.5 data set.

A least squares  sine fit to the the phased Ca II yields
an amplitude of 1.67 $\pm$ 0.25 m{\AA}
for the dates spanning 1986.5 -- 1992 (1.03 $\pm$  0.25 assuming a 906 day period).
For the dates covering  1981 -- 1986.5 the
amplitude of the $EW_{8662}$ variations is 0.48 $\pm$ 0.33 m{\AA}.
Clearly there are amplitude variations in the Ca II $\lambda$8662.

	We conclude that the periodic  variations in Ca II found by Walker
et al. (1992) are not long-lived and were only present during 1986.5--1992.
This period was clearly not present in our Ca II S-index measurements 
spanning 1998--2002. Since the RV
variations are unchanged during the entire time span 1981--2002 we do not
believe that the 906 day RV period is related to the cause of the Ca II 
$\lambda$8662\,{\AA} variability.

\subsection{The Spectral Classification of $\gamma$ Cep}

	One reason that Walker et al. (1992) favored rotational modulation
for the residual RV variations was the reclassification of $\gamma$ Cep
as a K0III. This was based on a visual comparison of the spectrum of
$\gamma$ Cep to other K giants (Bohlender et al. 1992). Recently, Fuhrmann
(2003) presented an extensive spectral analysis of nearby stars in the
galactic disk and halo. Included in this study was $\gamma$ Cep. 
Table~\ref{parameters} list the stellar properties derived by Fuhrmann.
Also listed is the {\it Hipparcos} distance ($d_{Hip}$) and 
distance determined spectroscopically ($d_{Sp}$) 
using the derived stellar parameters. These distances agree to within
3\% indicating that we can have some confidence in the results of the spectral
analysis. Fuhrmann's classification of K1IV for $\gamma$ Cep is
consistent with the sub-giant status for  $\eta$ Cep (K0IV) which has
a comparable effective temperature (4990 K), gravity (log $g$ = 3.4),
radius (4.14 $R_\odot$), and bolometric magnitude ($M_{bol}$ = 2.3).
The most current and best analysis of  $\gamma$ Cep supports this 
being a subgiant star.

\section{Searching for additional companions}

The long time-baseline of RV-monitoring of $\gamma$~Cep and the emergence of several 
multiple extrasolar planetary systems from Doppler surveys 
(e.g. Butler et al. 1999), 
encouraged
us to search for additional companions in this system. 
For this purpose we performed a period search within the RV-residuals, after subtracting both the binary 
orbital motion and the first planetary signal. 

Fig.~\ref{residper} shows the Lomb-Scargle periodogram 
of the RV-residuals in the period range of
$2$ to $7900$ days (frequency = 0.00013 -- 0.5 c\,d$^{-1}$). The strongest peak is found at $P\approx11$ days but its significance level is 
very low (false-alarm-probability (FAP) is $6.5\%$). The FAP-levels shown in the
figure were determined by $10,000$ runs of 
a bootstrap randomization scheme.

It is clear from this analysis that  
no additional periodic signal above the noise level is present in the RV-residuals of $\gamma$~Cep and 
we thus conclude that the 906-day period planet is the only giant planet in this system
evident in our data..

\section{Discussion }

	Precise stellar radial velocity measurements of $\gamma$~Cep
from 4 independent data 
sets spanning over 20 years show long-lived, low-amplitude RV variations
($K$ = 27.5 m\,s$^{-1}$) superimposed on the larger radial velocity variations
due to the reflex motion caused by a stellar companion. We interpret
the low amplitude, shorter period variations as due to the presence of
a planetary companion with $M$ $sin$\,$i$ = 1.7 $M_{Jupiter}$ and
an orbital semi-major axis, $a$ = 2.13 AU. Our conclusion that these 
short period variations are not due to rotation, pulsations,
 or changes in the convection pattern of the star is based on several
facts:

\begin{enumerate}
\item The 2.48 yr period has been present for over 20 years with
no changes in phase or  amplitude during this time.

\item No variations with this period are seen 
in the McDonald Ca II S-index measurements.

\item Spectral line bisector span and curvature measurements for $\gamma$~
Cep that are constant to less than 5 m\,s$^{-1}$ over an orbital cycle of
the planet.

\item Contemporaneous
{\it Hipparcos} photometry that is constant to less than 0.001 mag over an 
orbital cycle.

\item The periodic variations in the Ca II $\lambda$8662\,{\AA} found by Walker et al.
(1992) were only present during 1986.5--1992 and
are thus most likely not associated
with the residual RV variability observed for this star.

\end{enumerate}

	The CFHT Ca II data  do show evidence for long period variability.
Although this signal is weakly detected in the full data set, it is much
stronger in only the last half of the data set spanning
1986.5--1992, and completely absent  in the data from
1981 -- 1986.5. Furthermore, our S-index measurements made
during 1997 -- 2003 show no modulation.
If rotational modulation was responsible for the residual
RV variations then we would have seen RV amplitude changes. However,
the amplitude and phase for the residual RV variations remained
constant. It therefore seems unlikely that rotational modulation is the cause
of the 906 day period. Furthermore, the best fit period to the long
term Ca II variations during 1986.5 -- 1992 is 781 $\pm$ 116 days,
which is significantly less than the 906 day RV period. 

	Gamma Cep shows no Ca II modulation during  epochs centered on
1994 and 2000, and modulation (781 day period) during a time 
centered on 1989. This indicates a possible ``activity cycle'' period
of 10 -- 15 years, much longer than the presumed planet period.

	Our spectral line bisector measurements, which are constant,
 also support our conclusion
that we are not seeing rotational modulation. One could argue that the
residual RV variations are due to changes in the convection pattern
in the star and not due to magnetic structure (plage, spots). For instance,
if the ratio of areas of convective, hot rising cells  and intergranule,
sinking lanes changes with an activity cycle (in this case with a period
of $\sim$ 900 days), then the amount of convective blue (red) shift would
change periodically resulting in a measured RV signal. If the stellar magnetic
fields are strong enough to alter the convection pattern of the star, but
too weak to cause chromospheric structure, then we would see RV variations
without strong variations in the Ca II emission. However, in this case
{\it we should also see changes in the spectral line bisectors}. The
lack of significant variability  in the spectral 
line bisectors does not support the hypothesis of a changing convection 
pattern on the star, or a least changes that can influence the RV measurements.

	The inclination angle of the rotation axis can be
estimated by comparing the expected equatorial rotational velocity to the 
projected rotational velocity (v\,sin\,$i$), assuming 
that 781 days (period of the Ca II $\lambda$8662\,{\AA} variations) is 
the true rotation period of 
$\gamma$~Cep. Fuhrmann (2003) estimates a radius  $R$ = 4.66 $R_\odot$ which
gives an equatorial rotational velocity of 4.9 km\,s$^{-1}$. We have measured
a v\,sin\,$i$ $=$ 1.5 $\pm$ 1.0 km\,s$^{-1}$ which is consistent to the
value determined by Fuhrmann (2003). 
This yields a sin\,$i$ = 0.5 -- 0.1. 
Assuming that the orbital and rotation axis of the star is aligned,
the true mass for the planetary companion to $\gamma$~Cep is $\approx$ 
3--16 $M_{Jupiter}$. 

	Given that the 906-day RV variations are due to a sub-stellar
companion, the question arises whether this system is stable. After all,
both the planet and the stellar companion have modestly 
eccentric orbits.
Preliminary indications are that this system is indeed stable  (Dvorak
et al. 2003). (The study by Dvorak et al. was made prior to the publication
of our paper and it used preliminary orbital parameters published only
in conference abstracts. The stability analysis should be re-done using
the final parameters and error estimates we have presented here.) 

	The $\gamma$~Cep system presents a very interesting system
for the study of planet formation. Planets have been found around
host stars that are in a binary system, but these are widely 
separated pairs so the presence of the stellar companion may not have
an influence on the planet formation around one of the stars.
The $\gamma$~Cep binary is the shortest period binary for which
a planetary companion has been found in orbit around one of the
components. This implies that even in such a relatively close binary
the presence of the stellar companion does not hinder the 
process of planet formation.

	One final comment. RV surveys  have
had a stunning success  at finding planets in orbit around other
stars. Yet this is still an indirect detection method with the 
disadvantage that  various 
stellar phenomenon can mimic a planetary signal. RV
searches for extrasolar planets are, and should continue to be a careful
process with many candidates undergoing the painstaking process
of confirmation.  Several extrasolar planet candidates have a proved
to be due to stellar phenomena (Queloz et al. 2001; Butler et al. 2002) 
and more false detections
will undoubtedly be uncovered in the future. Likewise, radial 
velocity variations
that were initially attributed to stellar phenomena in the past may ultimately
prove to be due to a planetary companion, as is apparently the
case for $\gamma$~Cep. RV
searches for extrasolar planets must be accompanied by careful
studies of the host star and the measurement of a  various photometric and
spectroscopic (Ca II emission, line shapes) quantities. These measurements
are important
for they could have confirmed the planet around
$\gamma$~Cep seven years prior to the discovery of 51 Peg b.

This material is based upon work supported by the National Aeronautics
and Space Administration under Grant NAG5-9227 issued through the Office
of Space Science, and by National Science Foundation Grant AST9808980.
We thank Sebastian Els for his attempt to image the stellar secondary
using adaptive optics observations with the 4.2m-William Herschel telescope.

\begin{deluxetable}{lcc}
\tablecaption{McDonald Observatory Phase I radial velocities}
\tablewidth{0pt}
\tablehead{\colhead{JD-2400000.0}&\colhead{Radial Velocity}&\colhead{$\sigma$}\\
\colhead{ } & \colhead{[m\,s$^{-1}$]} & \colhead{[m\,s$^{-1}$]} }
\startdata
47368.9657 &  550.0 &    19.0  \\
47369.9300 &  570.8  &   19.0  \\
47369.9346 &  572.0 &    19.0  \\
47405.9338 &  500.8 &    19.0  \\
47430.7110 &  530.6 &    19.0  \\
47430.7141 &  520.5 &    19.0  \\
47459.7569 &  495.2 &    19.0  \\
47460.7559 &  488.8 &    19.0  \\
47495.7539 &  416.4 &    19.0  \\
47495.7597 &  416.3 &    19.0  \\
47496.7046 &  437.9 &    19.0  \\
47516.6665 &  440.0 &    19.0  \\
47517.6587 &  416.1 &    19.0  \\
47551.6057 &  397.0 &    19.0  \\
47551.6101 &  391.9 &    19.0  \\
47582.5740 &  359.1 &    19.0  \\
47696.9685 &  274.9 &    19.0  \\
47762.9403 &  220.3 &    19.0  \\
47785.9286 &  152.9 &    19.0  \\
47785.9319 &  157.6 &    19.0  \\
47786.8176 &  171.1 &    19.0  \\
47786.8210 & 166.3  &   190  \\
47813.7499 &  133.5 &    19.0  \\
47848.6803 &  102.9 &    19.0  \\
47879.6591 &  70.0  &   19.0  \\
47880.7112 &  67.7  &   19.0  \\
47895.6235 &  72.0  &   19.0  \\
48145.8981 &  -112.5&     19.0  \\
48145.9039 &  -112.2&     19.0  \\
48146.8217 &  -89.4 &    19.0  \\
48176.8457 &  -143.4&     19.0  \\
48176.8513 &  -153.7&     19.0  \\
48198.8661 &  -182.7&     19.0  \\
48227.7128 &  -178.0&     19.0  \\
48523.8192 &  -277.6&     19.0  \\
48523.8241 &  -278.0&     19.0  \\
48854.9066 &  -390.1 &    19.0  \\
48903.8517 &  -426.3 &    19.0  \\
49260.7901 &  -327.9 &    19.0  \\
49260.7946 &  -325.2 &    19.0  \\
49649.7368 &  -235.8   &  19.0  \\
49649.7390 &  -243.4  &   19.0  \\
49649.7411 &  -234.7 &    19.0      \\
\enddata
\label{phaseI}
\end{deluxetable}

\begin{deluxetable}{lcc}
\tablecaption{McDonald Observatory Phase II radial velocities}
\tablewidth{0pt}
\tablehead{\colhead{JD-2400000.0}&\colhead{Radial Velocity}&\colhead{$\sigma$}\\
\colhead{ } & \colhead{[m\,s$^{-1}$]} & \colhead{[m\,s$^{-1}$]} }
\startdata
48177.8696  & 35.39  & 33.09 \\
48177.8773  & 29.94 &  23.82 \\
48200.8311  & 27.72 &  13.42 \\
48224.7008  & 53.95 &  11.53 \\
48259.6020  & 42.85 &  13.09 \\
48484.8393  & -37.21&   14.00 \\
48524.8304  & -93.10 &  17.55 \\
48555.7939  & -105.15&   12.07 \\
48555.8038  & -116.01 &  9.08 \\
48607.7638  & -138.18 &  13.85 \\
48644.6465  & -137.70&  16.41 \\
48824.9547  & -204.50&  20.26 \\
48824.9609  & -192.72&  24.48 \\
48852.9733  & -209.83&   20.32 \\
48853.9149  & -212.64&   18.76 \\
48882.8273  & -205.27 &  12.96 \\
48901.7912  & -225.14  & 10.72 \\
48902.7764  & -235.24 &  18.78 \\
48943.7443  & -246.38  & 15.91 \\
48971.6571  & -238.61  & 18.23 \\
48973.6189  & -205.78  & 19.50 \\
49020.6401  & -218.97  & 13.72 \\
49220.9732  & -167.29  & 33.39 \\
49258.8484  & -132.89  & 17.28 \\
49286.7903  & -181.63  & 15.92 \\
49352.6804  & -119.99  & 18.05 \\
49380.5871  & -97.26  & 19.21 \\
49400.5656  & -116.95  & 17.10 \\
49587.8845  & -68.05  & 16.10 \\
49587.8907  & -61.08  & 17.55 \\
49615.8899  & -72.07  & 14.53 \\
49647.8128  & -73.35 & 10.00 \\
49670.7118  & -28.25 &  8.18 \\
49703.6783  & -35.81 &  12.4 \\
49734.5877  & -28.56  & 15.26 \\
49769.5639  & -22.20  & 11.46 \\
49917.9326  & 87.22  & 12.25 \\
49946.9247  & 103.39 &  36.52 \\
49963.8567  & 135.27 &  20.53 \\
49993.8304  & 127.05 & 16.89 \\
50093.6243  & 188.90 &  8.42 \\
50124.6006  & 222.54 &  8.77 \\
50292.8908  & 351.62 &  13.65 \\
50354.8007  & 389.66 &  9.85 \\
50409.7363  & 394.82 &  14.13 \\
50480.6530  & 420.67 &  15.16 \\
50700.8995  & 556.24 &  16.33 \\
50768.8008  & 575.22 &  14.58 \\
50834.6049  & 650.27 &  12.11 \\
\enddata
\label{phaseII}
\end{deluxetable}

\begin{deluxetable}{lcc}
\tablecaption{McDonald Observatory Phase III radial velocities}
\tablewidth{0pt}
\tablehead{\colhead{JD-2400000.0}&\colhead{Radial Velocity}&\colhead{$\sigma$}\\
\colhead{ } & \colhead{[m\,s$^{-1}$]} & \colhead{[m\,s$^{-1}$]} }
\startdata
51010.8960  & -583.71 &  7.55 \\
51010.9008  & -578.23 &  7.61 \\
51065.8469  & -555.03 &  7.60 \\
51152.6120  & -471.87 &  8.93 \\
51212.5924  & -420.20 & 8.78 \\
51212.5964  & -422.45 &  8.63 \\
51417.9144  & -328.56 &  7.93 \\
51451.8399  & -306.73 &  8.01 \\
51503.6547  & -290.01 &  7.93 \\
51503.6584  & -293.29 &  7.93 \\
51503.6617  & -294.87 &  7.76 \\
51530.7206  & -280.93 &  8.42 \\
51556.6279  & -266.48 &  8.48 \\
51750.9434  & -146.12 &  7.14 \\
51775.8747  & -114.43 &  7.88 \\
51811.8113  & -83.07 &  7.64 \\
51919.5884  & -13.88 &  7.89 \\
51946.7192  & 13.03  & 9.18 \\
52117.9545  & 131.98 &  7.77 \\
52221.8443  & 164.75 &  8.37 \\
52328.6118  & 194.18 &  7.82 \\
52472.9542  & 260.48 &  7.64 \\
52472.9575  & 258.52 &  8.06 \\
52473.9522  & 248.31 &  7.68 \\
52473.9553  & 246.10  & 7.77 \\
52492.8857  & 250.44 &  7.62 \\
52492.8876  & 241.13 &  7.43 \\
52493.8549  & 256.65 &  7.93 \\
52493.8567  & 244.26 &  7.66 \\
52494.9286  & 257.24 &  7.66 \\
52494.9302  & 264.18 &  7.76 \\
52495.9152  & 255.16 &  7.61 \\
52538.8445  & 284.82 &  8.07 \\
52538.8465  & 278.53 &  7.97 \\
52576.7817  & 306.18 &  7.90 \\
52599.6345  & 331.50  & 8.28 \\
52599.6370  & 314.88 &  8.29 \\
52621.7124  & 323.18 &  8.17 \\
52621.7148  & 324.99 &  8.41 \\
\enddata 
\label{phaseIII}
\end{deluxetable}

\begin{deluxetable}{lc}
\tablecaption{Velocity Offsets for the Datasets}
\tablewidth{0pt}
\tablehead{\colhead{Dataset}&\colhead{Velocity Offset}  \\
\colhead{ } & \colhead{[m\,s$^{-1}$]}  }
\startdata
CFHT & 1294.6 $\pm$ 108  \\
McD Phase I & 2232.3 $\pm$ 108 \\
McD Phase II & 2040.8 $\pm$ 110 \\
McD Phase III & 864.7 $\pm$ 108 \\
\enddata
\label{offsets}
\end{deluxetable}

\begin{deluxetable}{cll}
\tablecaption{Binary Orbital  Elements for $\gamma$~Cep }
\tablewidth{0pt}
\tablehead{
\colhead{Element}    &
 \colhead{This work} & \colhead{Griffin et al. (2002)}}
\startdata
Period  (days)       & 20750.6579 $\pm$  1568.6  & 24135 $\pm$ 349   \nl
T (JD)               & 248429.03 $\pm$ 27.0     & 248625 $\pm$ 210  \nl
Eccentricity         & 0.361 $\pm$ 0.023       & 0.389 $\pm$ 0.017   \nl
$\omega$ (deg)       & 158.76 $\pm$ 1.2        & 166 $\pm$ 7   \nl
K1 (km\,s$^{-1}$)     & 1.82   $\pm$ 0.049     & 2.04 $\pm$ 0.10   \nl
$f(m)$ (solar masses)& 0.0106 $\pm$ 0.0012    & 0.0166 $\pm$ 0.0025 \nl
Semi-major axis (AU) & 18.5 $\pm$ 1.1        & 20.3 $\pm$ 0.7\nl
Reduced $\chi^2$ (without planet) & 4.36      &  \nl
\enddata
\label{binorbit}
\end{deluxetable}

\begin{deluxetable}{cl}
\tablecaption{Orbital  Elements for Planet around $\gamma$~Cep }
\tablewidth{0pt}
\tablehead{
\colhead{Element}    &
 \colhead{Value} }
\startdata
Period  (days)       & 905.574 $\pm$  3.08     \nl
T (JD)               & 253121.925  $\pm$ 66.9     \nl
Eccentricity         & 0.12   $\pm$ 0.05     \nl
$\omega$ (deg)       & 49.6 $\pm$ 25.6       \nl
K1 (m\,s$^{-1}$)    & 27.50  $\pm$ 1.5        \nl
$f(m)$ (solar masses)& (1.90 $\pm$ 0.3)  $\times$10$^{-9}$   \nl
Semi-major axis (AU) & 2.13 $\pm$ 0.05       \nl
Reduced $\chi^2$  & 1.47        \nl
$\sigma_{CFHT}$ (m\,s$^{-1}$) & 15.3  \nl
$\sigma_{PhaseI}$ (m\,s$^{-1}$) & 17.4  \nl
$\sigma_{PhaseII}$ (m\,s$^{-1}$) & 15.8  \nl
$\sigma_{PhaseIII}$ (m\,s$^{-1}$) & 8.2  \nl
\enddata
\label{planorbit}
\end{deluxetable}

\begin{deluxetable}{cl}
\tablecaption{Stellar Parameters (Fuhrmann 2003)} 
\tablewidth{0pt}
\tablehead{
\colhead{Parameter}    & \colhead{Value} }
\startdata
$T_{eff}$ (days)    &  4888 K\nl
log $g$             &  3.33      \nl
$[Fe/H]$              & +0.18 \nl
$M_{bol}$           & 2.14 \nl
Mass               & 1.59 $M_\odot$     \nl
Radius             & 4.66 $R_\odot$     \nl
$d_{Hip}$            &  13.79 pcs \nl
$d_{Sp}$             &  13.39 pcs \nl
\enddata
\label{parameters}
\end{deluxetable}
\clearpage

\begin{figure}
\plotone{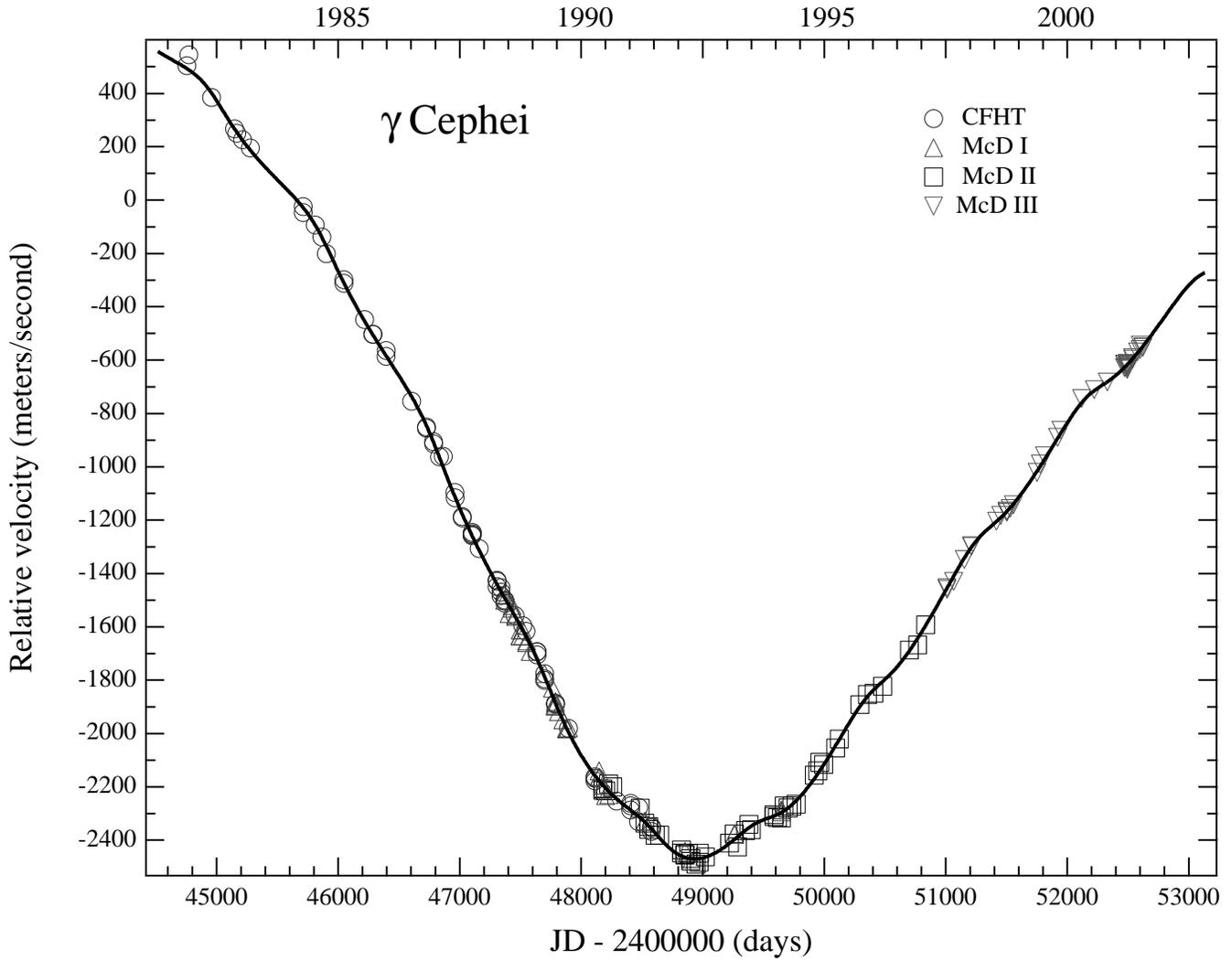}
\figcaption[ftall.ps]{The combined (planet + stellar) orbital solution
to all RV data sets for $\gamma$~Cep. Circles represent the
CFHT measurements of Walker et al. (1992). All other symbols
are for data taken at McDonald Observatory. Triangles represent 
measurements using telluric O$_2$ as the reference (Phase I), squares are
for measurements using an iodine absorption cell (Phase II), and inverted
triangles are also I$_2$ measurements, but with the large wavelength
coverage 2dcoud{\'e} spectrograph (Phase III).
\label{orbit}}
\end{figure}

\epsscale{0.7}
\begin{figure}
\plotone{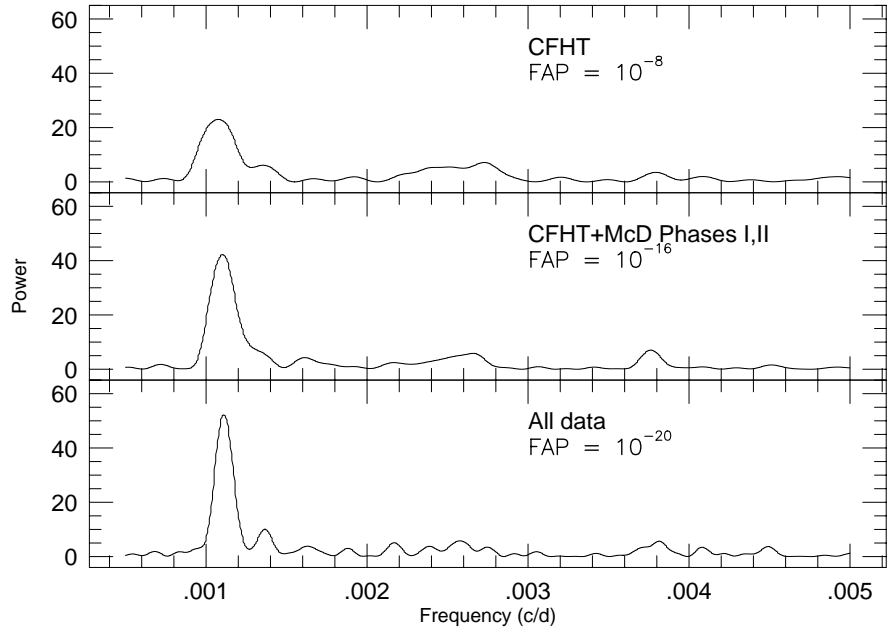}
\figcaption[ftall.ps]{The Lomb-Scargle periodogram of the combined
RV measurements for $\gamma$~Cep after removal of the RV variations
due to the stellar companion. (Top) The CFHT data alone. (Middle) The
periodogram of the 
combined CHFT + McDonald Phases I,II data. 
(Bottom) Periodogram of all measurements. The false alarm
probability (FAP) for the peak in each periodogram is shown
in each panel. This was computed using the equation given in
Scargle (1982).
\label{periodogram}}
\end{figure}

\begin{figure}
\plotone{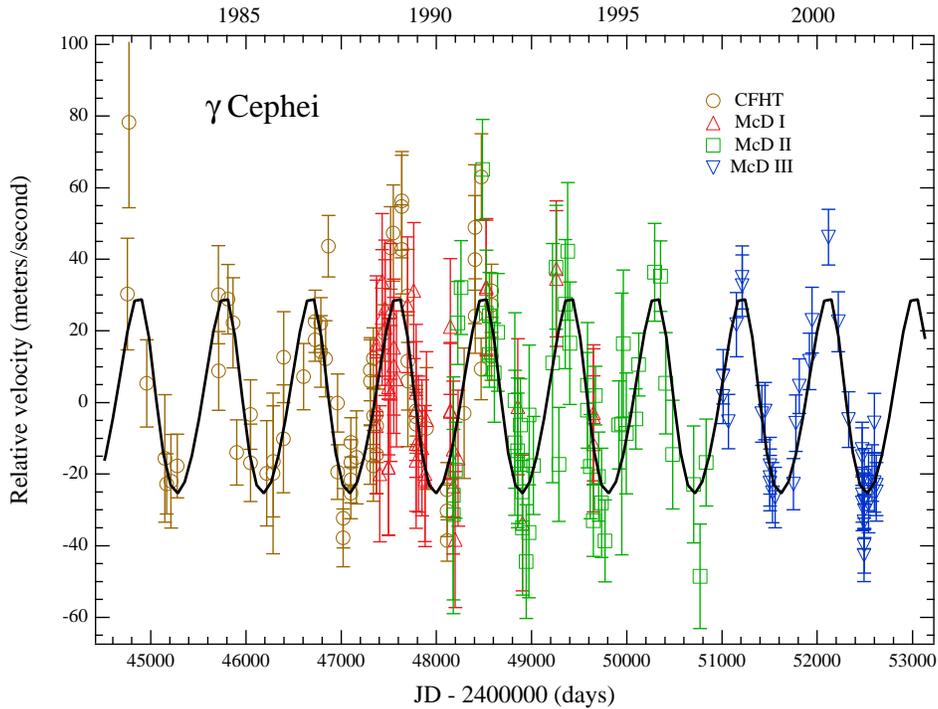}
\figcaption[residuals.os]{The orbital solution for the planet (line) and
the residual velocity measurements of the 4 data sets after subtracting
the contribution due to the binary companion (points).
\label{residuals}}
\end{figure}

\begin{figure}
\plotone{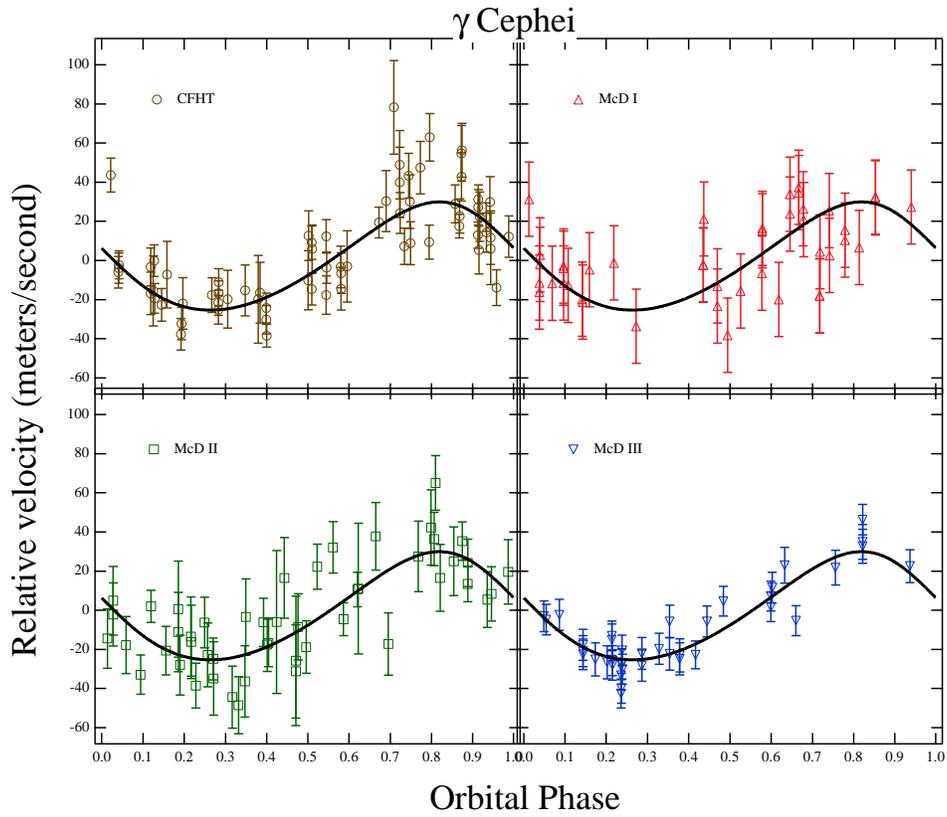}
\figcaption[ftall.ps]{The phased residual RV measurements of CFHT, 
and McDonald Phase I--III compared to the planet orbital solution (line).
\label{phase}}
\end{figure}

\begin{figure}
\plotone{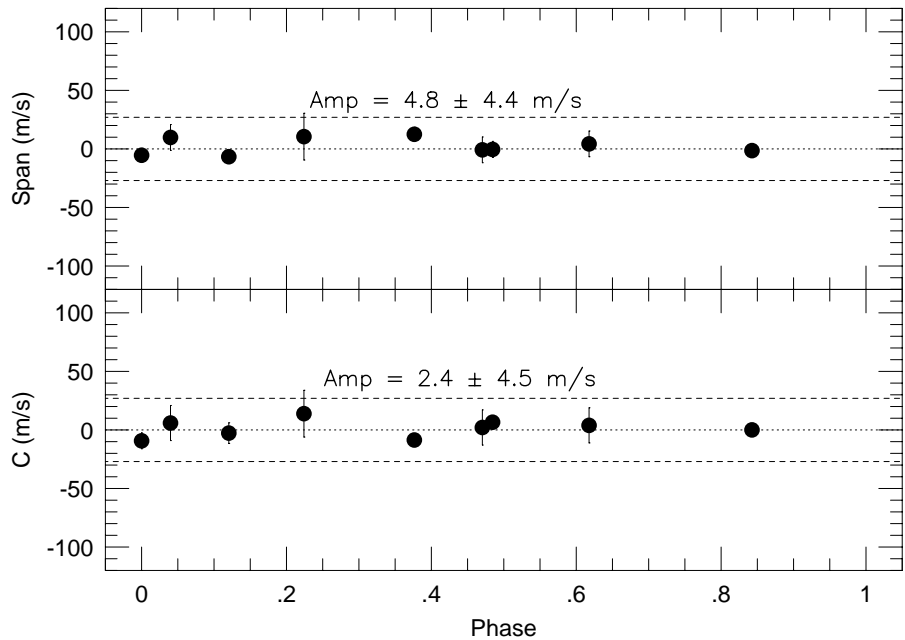}
\figcaption[]{The mean bisector span measurements (top)
and the mean bisector curvature (bottom)
measurements for $\gamma$~Cep phased to the planet orbital period.
The dotted line marks zero value and the dashed line represents
the extreme values of the radial velocity variations due to the
planetary companion. 
\label{bisector}}
\end{figure}

\begin{figure}
\plotone{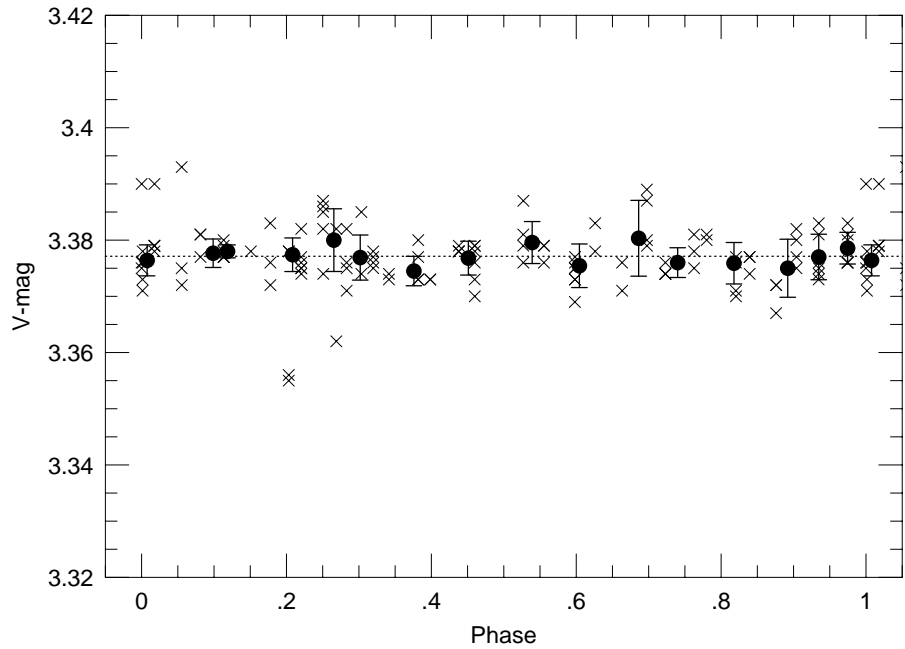}
\figcaption[]{
The {\it Hipparcos} photometry for $\gamma$~Cep over 1989.9 -- 1993.17 phased
to the planet orbital period. The crosses represent the individual
measurements and the solid points phased-binned averages. The error
bars indicate the rms scatter of values used in computing the
binned average. 
\label{photometry}}
\end{figure}

\begin{figure}
\plotone{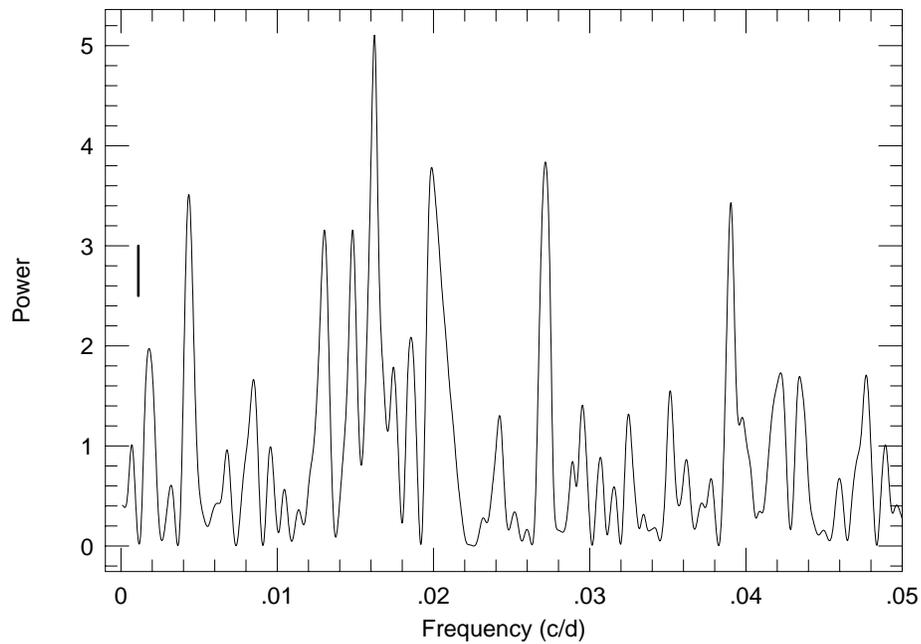}
\figcaption[]{The Lomb-Scargle periodogram of the {\it Hipparcos} (daily averages)
photometry. The vertical line marks the orbital frequency of the
planet.
\label{photft}}
\end{figure}

\begin{figure}
\plotone{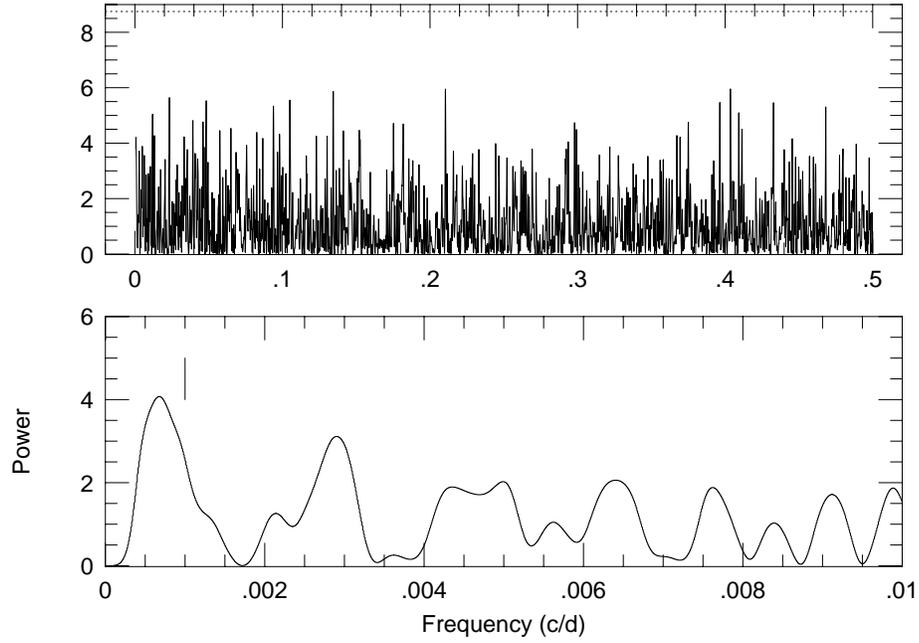}
\figcaption[]{
Periodogram of the S-index measurements using the McDonald Phase III data. The vertical
line in the lower, expanded scale plot marks the location of the orbital frequency 
of the planet. The horizontal dashed line in the top panel indicates
a false alarm probability of 1\%.
\label{sindexft}}
\end{figure}

\begin{figure}
\plotone{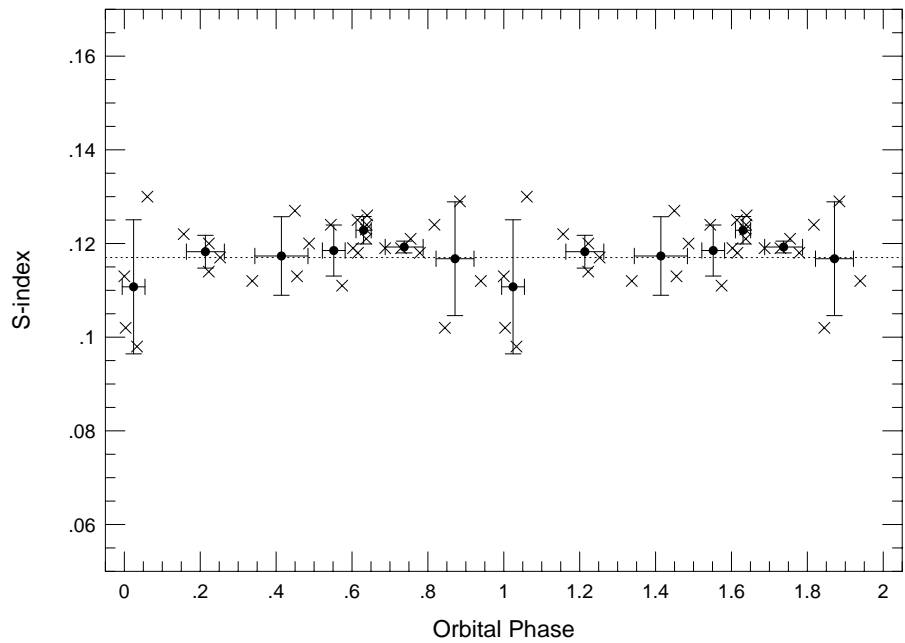}
\figcaption[]{
The S-index measurements from  the McDonald Phase III data phased to the orbital
period of the planet. Crosses represent the individual measurements,
solid points phased-binned averages. The error bars represent the rms
scatter of the data used for the binned averages. 
\label{sindex}}
\end{figure}

\begin{figure}
\plotone{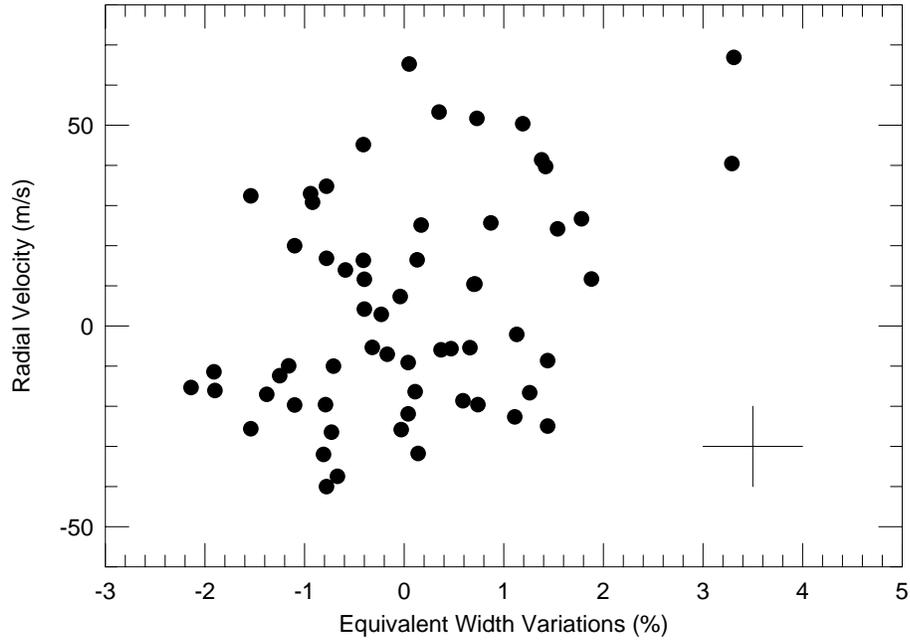}
\figcaption[]{The correlation between the RV and the
Ca II $\lambda$8662\,{\AA} equivalent width (top) and the RV and the S-index
measurements. A typical error bar is shown in the lower right of the figure.
\label{cacorr}}
\end{figure}

\begin{figure}
\plotone{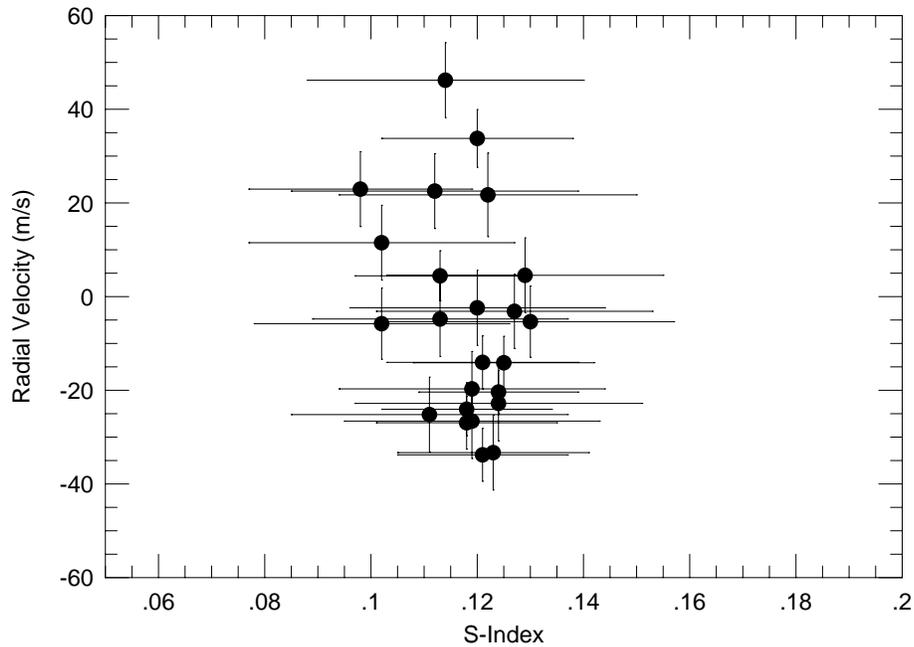}
\figcaption[]{The correlation between the RV and the McDonald 
Ca II S-index measurements 
measurements.
\label{mcdcorr}}
\end{figure}

\begin{figure}
\plotone{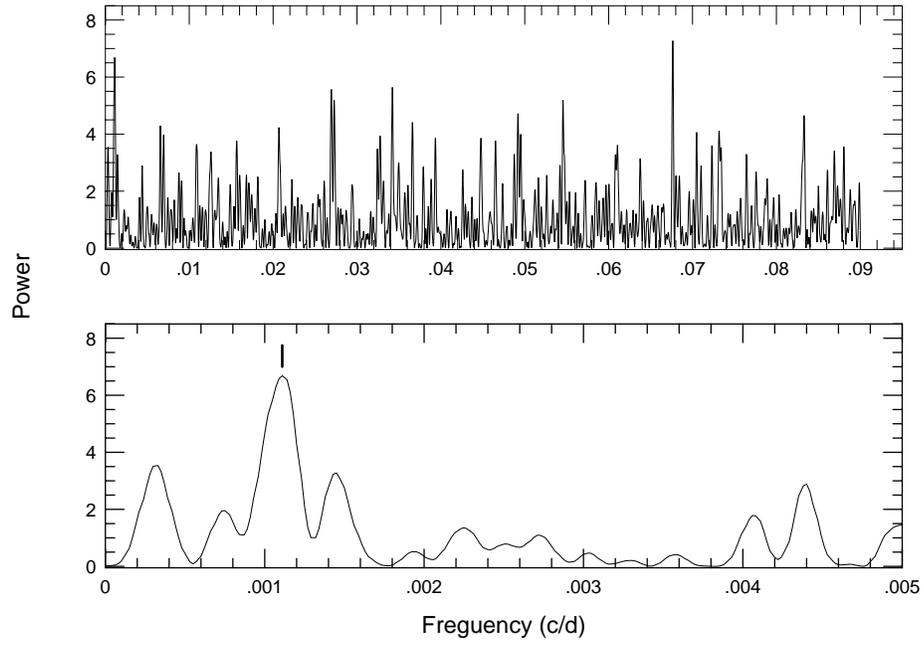}
\figcaption[]{The periodogram of the CFHT Ca II $\lambda$8662\,{\AA}
equivalent width measurements. The lower panel is an
expanded scale near the planet orbital frequency
indicated by the vertical line.
\label{cfhtcaperiod}}
\end{figure}

\begin{figure}
\plotone{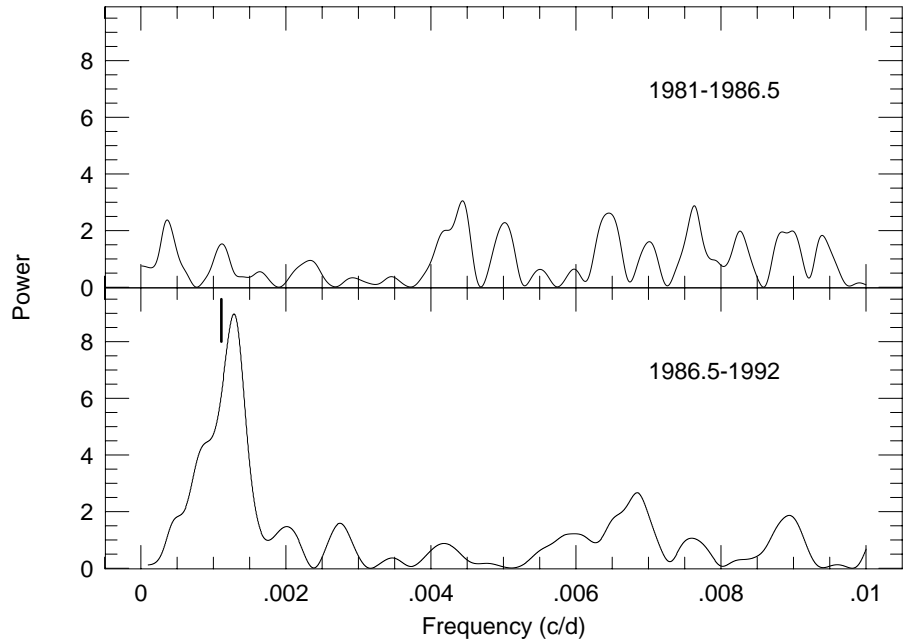}
\figcaption[]{(Top)  The Lomb-Scargle
periodogram for the CHFT Ca II $\lambda$8662\,{\AA} 
measurements over the time span 1981 -- 1986.5.
(Bottom) The same for the time span  1986.5 -- 1992.
The vertical line marks the orbital frequency of the planet.
\label{casplit}}
\end{figure}

\begin{figure}
\plotone{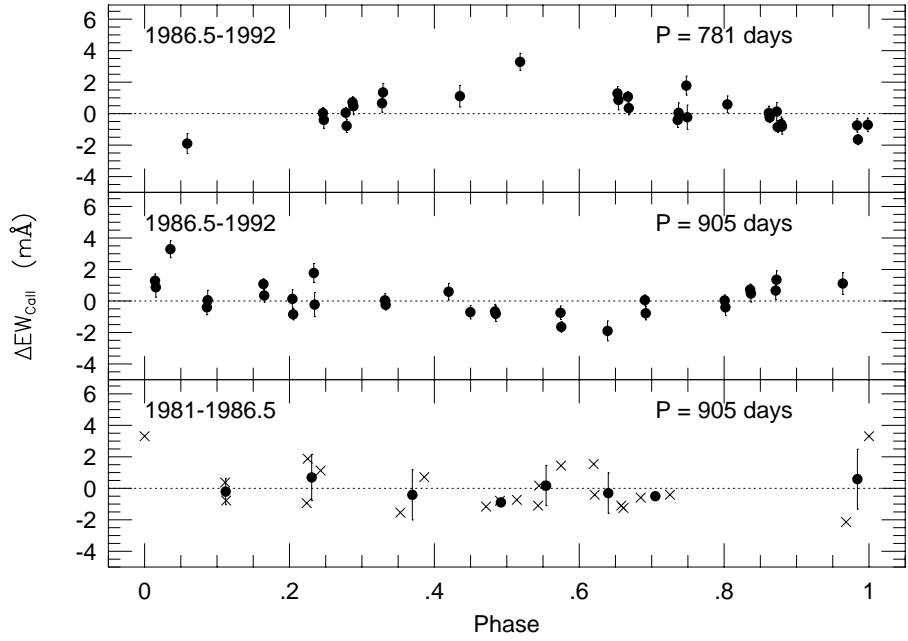}
\figcaption[]{The 1986.5 -- 1992 CFHT Ca II
measurements phased to the period found in the periodogram
analysis (top) and the planet period (middle). The
bottom panel are the CFHT Ca II equivalent width variations
from 1981-1986.5 phased to the planet period.
\label{2phase}}
\end{figure}

\begin{figure}
\plotone{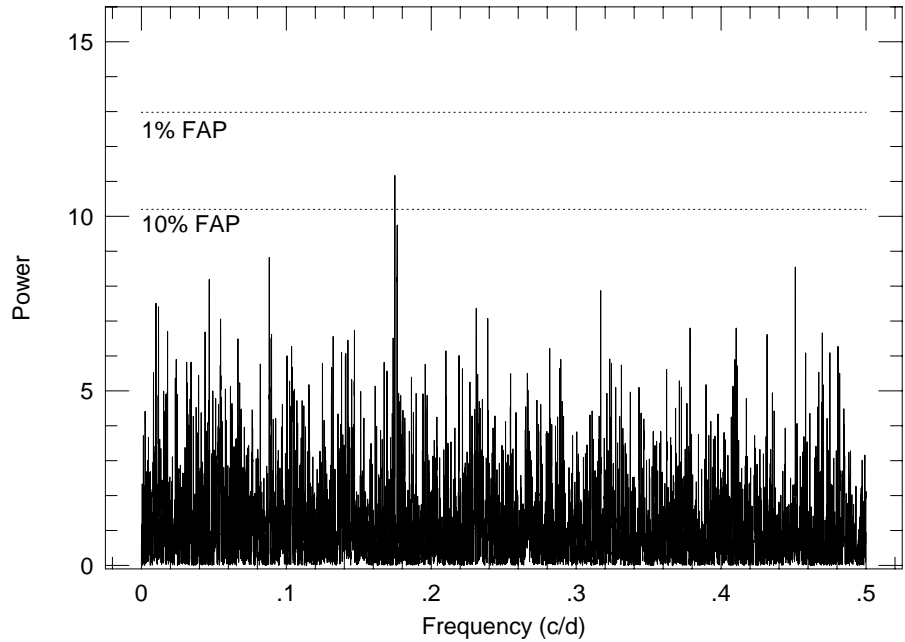}
\figcaption[]{Lomb-Scargle periodogram of the RV-residuals of $\gamma$~Cep
after subtracting the binary and the planetary signal. The horizontal dashed lines
show confidence levels with $10\%$ and $1\%$ false-alarm-probability (FAP). Obviously,
no additional periodic signal above the noise level is present in the data. 
\label{residper}}
\end{figure}

\end{document}